\documentclass{mem}
\usepackage{natbib}
\usepackage{txfonts}
\usepackage{balance}
\usepackage{txfonts}
\usepackage{graphicx}
\usepackage[a4paper]{hyperref}
\idline{75}{282}
\begin{document}
\def\teff{$T\rm_{eff }$}
\def\kms{$\mathrm {km s}^{-1}$}

\title{
Models of Comptonization}

   \subtitle{}

\author{P.O. Petrucci
          }

  \offprints{P. O. Petrucci}

\institute{
Laboratoire d'Astrophysique de Grenoble,
UJF/CNRS,
414 rue de la Piscine,
38041 Grenoble Cedex 9,
FRANCE\\
\email{pierre-olivier.petrucci@obs.ujf-grenoble.fr}
}

\authorrunning{Petrucci }

\titlerunning{Models of Comptonization}

\abstract{After a rapid introduction about the models of comptonization, we present some simulations that underlines the expected capabilities of Simbol-X to constrain the presence of this process in objects like AGNs or XRB
}
\maketitle{}

\section{The Comptonization process}
\begin{figure}[htbp]
\begin{center}
\includegraphics[width=\columnwidth]{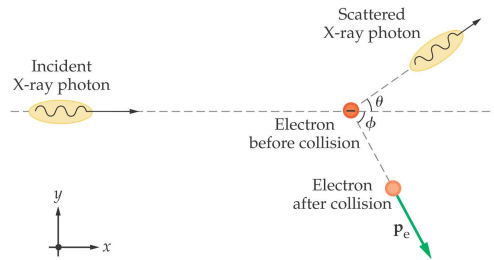}\\
\caption{{ A photon of energy E  comes in from the left, collides with a target (usually an electron) at rest, and a new photon of energy E'  emerges at an angle $\theta$.}}
\label{compton}
\end{center}
\end{figure}

The Compton effect was diskovered by A.H. Compton in 1923 and corresponds to the
gain or loss of energy of a photon when it interacts with matter (usually electrons).
For an electron at rest, the photon loss of energy is of the order
\begin{eqnarray*}
\Delta E &=& E'-E\\
&\simeq& -\frac{E^2}{m_ec^2}(1-\cos\theta)
\end{eqnarray*}
where $\theta$ is the photon scattering angle (see Fig. \ref{compton}).

\noindent For a on-stationnary electron part of the electron energy can be taken away by the photons (i.e. $\Delta E > 0$). This corresponds to the inverse Compton process.
\begin{figure*}[t!]
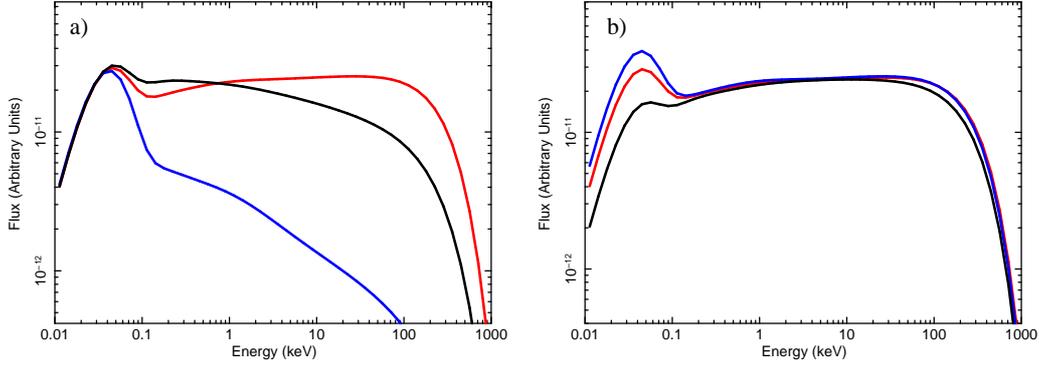

\begin{center}
\begin{tabular}{cc}
\includegraphics[height=\columnwidth,angle=-90]{petrucci_f2.ps}&
\includegraphics[height=\columnwidth,angle=-90]{petrucci_f3.ps}\\
\end{tabular}
{\begin{minipage}{2cm}
	\vspace*{-8.7cm}\hspace*{-4.8cm}
	a)
	\hspace*{6.7cm}
	b)
    \end{minipage}}
    \caption{{{\bf a)} Thermal comptonization spectra for the same plasma temperature and optical depth but different geometry: blue: cylindrical, red: slab, black: spherical. {\bf b)} We fix the plasma temperature but change its optical depth in order to have roughly the same spectral index in the 2-10 keV band. This exemplifies the "geometrical" degeneracy of the thermal comptonization spectrum.}}
\label{comp}
\end{center}
\end{figure*}

\subsection{Thermal Comptonization}
Thermal comptonization corresponds to the case where seed photons (the "cold" phase) are comptonized by a thermal plasma (the "hot" phase) of electron. This thermal plasma is characterized by a temperature $T_e$ and an optical depth $\tau$.  In this case, the mean relative energy gain per collision $\displaystyle\frac{\Delta E}{E}$ and the mean number of scatterings can be easily computed (e.g. \citep{ryb79}):
\begin{eqnarray*}
\frac{\Delta E}{E} &\simeq& \left( \frac{4kT_e}{m_ec^2}\right ) + 16\left( \frac{kT_e}{m_ec^2}\right )^2 \mbox{for}\  E\ll kT_e\\
\end{eqnarray*}
and
\begin{equation}
N\simeq (\tau+\tau^2)\nonumber
\end{equation}
Then we define the Compton parameter $\displaystyle y~=~\frac{\Delta E}{E}N$. Large values of $y$ means that the Comptonization process is efficient and modifies noticeably the seed spectrum.

\subsubsection{Thermal Comptonization spectrum}

While thermal comptonization spectra have been computed since more than
two decades, important effects like the anisotropy of the soft photon
field were precisely taken into account in the beginning of the 90's by
the pioneering works of \cite{haa93,ste95} and \cite{pou96}. These
effects appear far from being negligible. Noticeably, for a given set of plasma temperature and optical depth, the spectral shape
is significantly different for different disk-corona geometries (cf. Fig. \ref{comp}a) but also
for different viewing angle, this last dependence being crucial for a
correct interpretation of X-ray spectra. These effects
underline the differences between realistic thermal comptonization
spectra and the cut-off power law approximation generally used to mimic
them. Moreover, the relation between the photon index $\Gamma$ and the plasma temperature and optical depth can be strongly degenerated ("spectral" degeneracy) , different couples ($\tau, T_e$) giving, for the same disk-corona geometry, the same $\Gamma$. A "geometrical" degeneracy also exists and is exemplifies in Fig. \ref{comp}b where the same high energy spectrum is reproduced with different set of parameters and with different geometries. These degeneracies complicate the fitting procedure  and high S/N data over broad band energy intervals are required to break them.

\subsubsection{Radiative Balance}
\begin{figure}[b!]
\begin{center}
\includegraphics[width=\columnwidth]{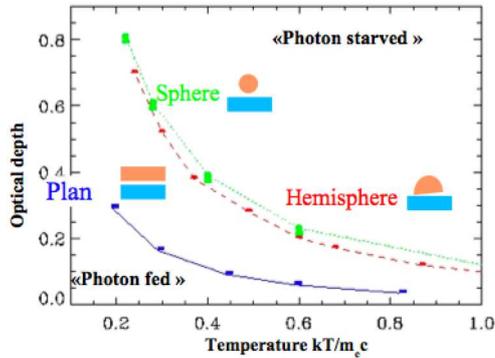}
\caption{{Different theoretical relationship, corresponding to different disk-corona geometry, between the corona temperature and optical depth in the case of energy balance between the hot and cold phases.}}
\label{thetavstau}
\end{center}
\end{figure}

In a disk-corona system, the Comptonizing region and the
source of soft photons are {\it coupled}, as the optically thick disk
necessarily reprocesses and reemits part of the Comptonized flux as soft
photons which are the seeds for Comptonization.  The system must then
satisfy equilibrium energy balance equations, which depend on {\it
geometry} and on the ratio of direct heating of the disk to that of the
corona. In the limiting case of a "passive" disk (i.e. non intrinsically radiative, it radiates what it absorbs from the hot phase), the amplification of
the Comptonization process, determined by the Compton parameter
$y$, is fixed by geometry only \citep{haa91,ste95}. Therefore, if the corona is in energy balance, the
temperature and optical depth must satisfy a relation which can be
computed for different geometries of the disk+corona configuration (cf. Fig. \ref{thetavstau}). It is then theoretically possible
to constrain the geometry of the system and verify the selfconsistency of
the model, provided that the plasma temperature  and optical depth $T_e$ and $\tau$ are known with sufficient
precision.

\subsection{Non-thermal Comptonization}
 In the case of non-thermal particles the energy transfert between the elctron and the photon is very efficient. For an electron with a Lorentz factor $\gamma$ it is of the order: 
 \begin{equation}
\Delta E\simeq \gamma^2 E
\end{equation}
Consequently the comptonization of a monoenergetic seed photon field by a non-thermal distribution of electrons $n(\gamma)\propto\gamma^{-s}$ produces a non-thermal spectrum $F(\nu)\propto\nu^{-\frac{s-1}{2}}$ (e.g. \citealt{ryb79}).

\section{What can we expect with SIMBOL-X?}
The comptonization process play a role in all SIMBOL-X science cases,
\begin{itemize}
\item AGNs (thermal comptonization in Seyfert galaxies, non-thermal comptonization in Blazars)
\item  X-ray binaries (thermal comptonization in the hard state, non-thermal(?) comptonization in the Intermediate and Soft states)
\item X-ray background
\item Galaxy clusters
\item Supernovae remnants
\item GRBs
\end{itemize} 
We detailed below some simulations of comptonization spectra showing the advances that we can expect with Simbol-X.  

\begin{figure}[b]
\begin{center}
\includegraphics[width=\columnwidth]{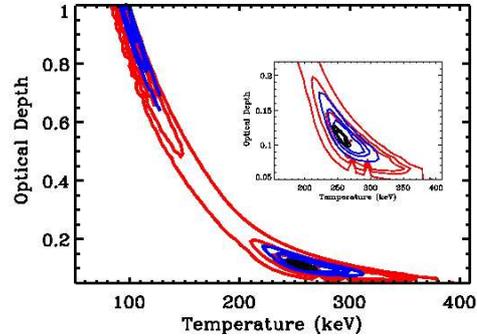}
\caption{{Contour plots $\tau$-$kT_e$ for simulations of different exposures (red: 1ks, blue: 5 ks and black: 50 ks). The simulated data correspond to a thermal comptonization spectrum with $L_{2-10 keV} = 10^{-11}$ erg.s$-1$.cm$-2$,  $kT_e =$  250 keV, $\tau =$ 0. Only the long exposure simulation enable to break the $\tau$-$kT_e$ degeneracy characteristic of thermal comptonization spectra.}}
\label{simu5548}
\end{center}
\end{figure}

\begin{figure*}[t]
\begin{center}
\begin{tabular}{cc}
\includegraphics[width=\columnwidth]{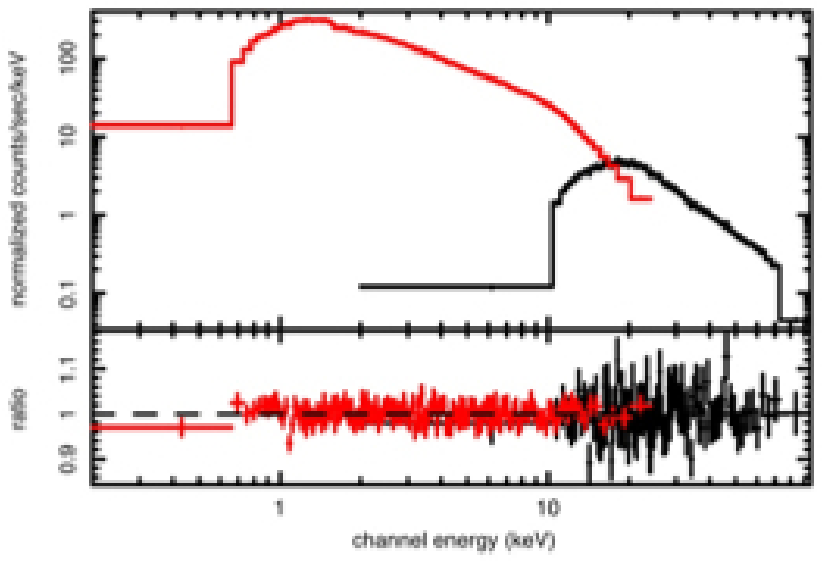}&\includegraphics[width=\columnwidth]{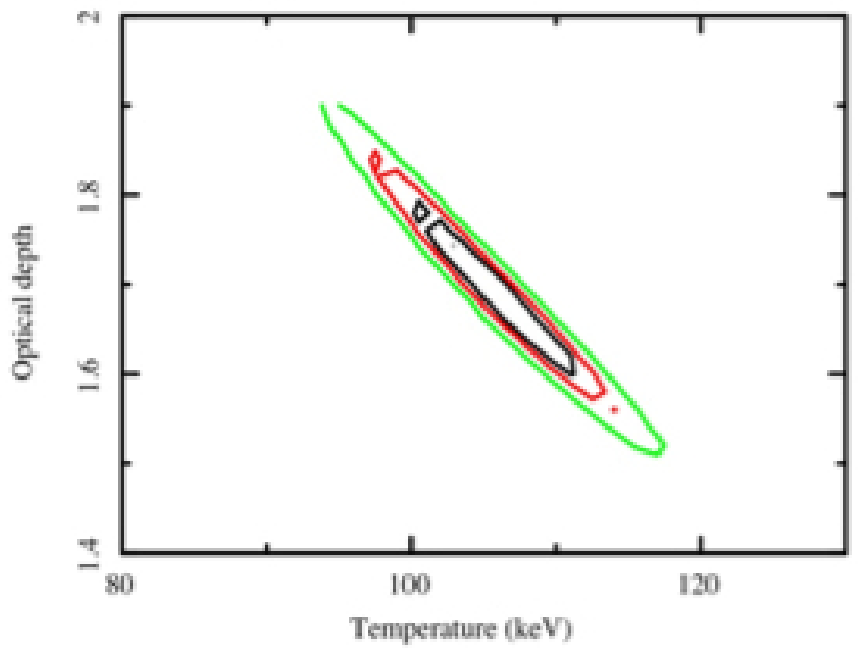}
\end{tabular}
\caption{{Left: 500 sec. simulated Simbol-X data of Cyg X-1 assuming $L_{2-10 keV}$ = 10$^{-9}$ erg.s$^{-1}$.cm$^{-2}$ $kT_e = 100$ keV, $\tau = 1.7$ and $R = 0.3$. Right: the corresponding  $\tau$-$kT_e$ contour plot.}}
\label{simucyg1}
\end{center}
\end{figure*}

\subsection{Simu. N$^\circ$ 1: the case of a Seyfert galaxy}
We simulate the spectrum of a Seyfert galaxy (NGC 5548) with different exposure time assuming $L_{2-10 keV} = 10^{-11}$ erg.s$-1$.cm$-2$,  $kT_e =$  250 keV, $\tau =$ 0.1 and $R = 1$. We assume a slab geometry. Then we fit the faked data with the model used for the simulations. The corresponding contour plots $\tau$-$kT_e$ are plotted in  Fig. \ref{simu5548}. If, for small exposure time (1 ks, red contour plot) the contour is very elongated, this "spectral" degeneracy is broken for exposures of a few tens of ks (see the black contours that correspond to 50 ks).  However we underline the fact that the real data generally show  the presence of complex reflection/absorption features that can strongly limit the data analysis and the precise study of the underlying continuum.

It is worth noting that a few tens of ks corresponds to the orbiting timescale at a few Schwarzschild radii around a $10^8$ solar masses black hole. It means that it should be possible with Simbol-X to follow the time evolution of the temperature and optical deth of the corona on a dynamical timescale.
However simulations show that breaking the ÒgeometricalÓ  degeneracy will require very long ($>$ 100 ks) exposure with Simbol-X.

\subsection{Simu. N$^\circ$2: the case of a microquasar}
We simulate the X-ray spectrum of the microquasar Cyg X-1, with $L_{2-10 keV}$ = 10$^{-9}$ erg.s$^{-1}$.cm$^{-2}$ $kT_e = 100$ keV, $\tau = 1.7$ and $R = 0.3$. Good constraints of the spectral parameters can be obtained in a few hundred of seconds (cf. Fig. \ref{simucyg1}).

\subsection{Simu. N$^\circ$3: the case of a blazar}
\begin{figure*}[htbp]
\begin{center}
\includegraphics[width=\textwidth]{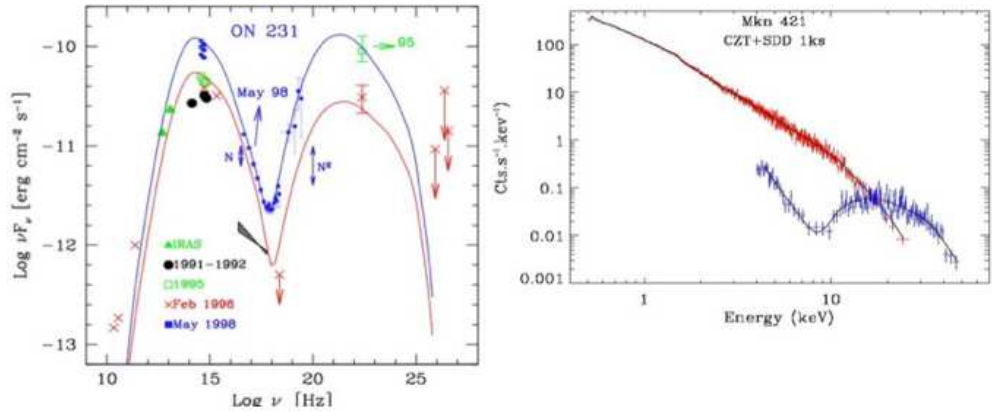}
\caption{{Left: Typical Synchrotron Self-Compton spectrum of a blazar. Right: Simbol X simulation of the blazar Mkn 421 in 1ks.}}
\label{simublazar}
\end{center}
\end{figure*}

We also simulate the X-ray spectrim of the blazar  Mkn 421 (cf. Fig. \ref{simublazar}). The spectrum is very well determine in 1 ks ! But in order to have constrains on the Synchrotron Self-Compton process  multi-$\lambda$ observations (radio, Optical, IR, X and $\gamma$) are needed.

\section{Conclusions}

\begin{itemize}
\item Simbol-X should bring strong constrains on comptonization spectra on dynamical time scale for AGNs, and on very short time scale in XrBs.
\item However this  can be complicated by the presence of complex absorption/emission features
\item The broadest energy range is needed and multi-wavelength observations are recommended (CTA, GLAST, HERSCHEL, ALMA, LOWFAR, WSO-UV, ...). 
\end{itemize}

\begin{acknowledgements}
I am very grateful to the CEA/DSM/DAPNIA/SAp for its financial support. 
\end{acknowledgements}


\end{document}